\documentclass[aps,showpacs,amsmath,amssymb]{revtex4}

\usepackage{latexsym}
\usepackage{graphicx}
\usepackage{dcolumn}
\usepackage{bm}
\usepackage{hyperref}

\newcommand{\beq}{\begin{equation}}
\newcommand{\eeq}{\end{equation}}
\newcommand{\beqa}{\begin{eqnarray}}
\newcommand{\eeqa}{\end{eqnarray}}
\newcommand{\beqar}{\begin{eqnarray*}}
\newcommand{\eeqar}{\end{eqnarray*}}
\newcommand{\bra}[1]{\mbox{$\left\langle{#1}\right|$}}
\newcommand{\ket}[1]{\mbox{$\left|{#1}\right\rangle$}}

\def\I{{\rm i}}
\def\d{{\rm d}}
\def\e{{\rm e}}

\newcounter{saveeqn}

\newcommand{\g}{\gamma}
\newcommand{\dif}[1]{\frac{\mathrm{d}}{\mathrm{d}#1}}
\newcommand{\Dif}[2]{\frac{\mathrm{d}^{#2}}{\mathrm{d}#1^{#2}}}

\begin{document}

\title{Quantum mechanics in the general quantum systems (V): Hamiltonian
eigenvalues}
\author{Zhou Li and An Min Wang}
\email{anmwang@ustc.edu.cn}
\altaffiliation{These authors made equal
contribution to
the paper.}

\affiliation{%
Quantum Theory Group, Department of Modern Physics, University of
Science and Technology of China, Hefei, 230026, P.R.China
}%
\homepage{http://qtg.ustc.edu.cn}%


\begin{abstract}
We derive out a complete series expression of Hamiltonian
eigenvalues without any approximation and cut in the general quantum
systems based on Wang's formal framework \cite{wang1}. In
particular, we then propose a calculating approach of eigenvalues of
arbitrary Hamiltonian via solving an algebra equation satisfied by a
kernal function, which involves the contributions from all order
perturbations. In order to verify the validity of our expressions
and reveal the power of our approach, we calculate the ground state
energy of a quartic anharmonic oscillator and have obtained good
enough results comparing with the known one.
\end{abstract}


\maketitle

\section{Introduction}
To determine the Hamiltonian eigenvalues is a basic problem in
quantum mechanics. As is well known, the exact solution of the
Schr\"odinger equation can be obtained only in some special cases,
i.e., for several elementary systems like the hydrogen atom, the
$H_2^+$ molecule, or the harmonic oscillator. In the majority of
cases, approximation techniques have to be employed in calculation
of the Hamiltonian eigenvalues in the general quantum systems, and
the precision and computability become focus.

Perturbation theory is one of the few principal methods of
approximating solutions to eigenvalue problems in quantum mechanics.
The formalism of Rayleigh-Schr\"odinger perturbation expansion
expresses an eigenvalue as a formal power series of the coupling
constant $\lambda$: \beq\label{pee} E_{\lambda}^{\rm
total}=\sum_{m=0}^{\infty}c_m \lambda^m\eeq  At the $4$th order
perturbation level, the explicit form of $c_4$ has appeared a little
complicated. Nevertheless, for a given $n$th order perturbation, one
can obtain, in principle, the form of $c_n$ according to Kato's
\cite{Kato} or Bloch's \cite{Bloch} formal expression.

The normal perturbation theory frequently meets two problems: one is
about whether the power series is convergent in some neighborhood of
$\lambda=0$ or it is only asymptotic as $\lambda \to 0$; another is
about the strong coupling regime. While the problems of strong
coupling is usually overcome with various kinds of renormalization
techniques, for example, the renormalization scheme recently used by
\v{C}\'i\v{z}ek and Vrscay \cite{renormalization1,renormalization2},
summation techniques are employed to give a divergent perturbation
series any meaning beyond a mere formal expansion \cite{summation}.

In our point of view, these problems closely connects with the whole
and deep knowledge about the expansion (\ref{pee}) because the
perturbation series is inherently multiple in the general quantum
systems. In mathematics, a reasonable rearrangement of a multiple
series is often significant when a cut approximation needs to be
introduced. Consequently, we wonder whether the expansion as a power
series of the coupling constant is a unique choice, and whether a
more explicit expression of total Hamiltonian eigenvalues exists.
Moreover, we would like to find a systematical and new approach for
the calculation of Hamiltonian eigenvalues in the general quantum
systems.

More than three year ago, An Min Wang, one of authors in this
article, presented his research on a formal framework of quantum
mechanics in the general quantum systems \cite{wang1,wang2,wang3}
and made a conjecture about the total Hamiltonian eigenvalues. Just
based on Wang's works, a complete series expression of Hamiltonian
eigenvalues in a general quantum system without any approximation
and cut is explicitly obtained by using some skills in mathematics
and physics. This expression is simply not a power series of
perturbed parameter, but a series of power of a kernal function as
well as its derivatives that involves the contributions from all
order perturbation. No cut and approximation are introduced, and the
general term is given out. In special, our kernal function as well
as its derivative is equal to $\mathcal{O}(\lambda^2)$. It implies
that our series is obviously improved in its approximation content
not only involving the higher order contributions but also being
more suitable to cut upto some given $n$th term since its decrease
is more rapidly than the series in the normal theory. Of course, if
expanding our expression of the total Hamiltonian according to the
perturbed parameter, our expression is consistent with one in the
normal perturbation theory. However, our expression really shows the
physics nature because it directly comes from the law of quantum
dynamics, but not a transcendent input -- expansion according to the
power of perturbed parameter which covers up a fact that
perturbation series is inherently multiple in the general quantum
systems. Moreover, it will be seen that the new conclusion can be
obtained by our expression.

It is worthy to point out that it is inevitable and acceptable that
the complete expression of Hamiltonian eigenvalues, just like ours,
is an infinite series, because that a general quantum system has
usually no a compact solution without any approximation. More
importantly and interestingly, from our expression we further
present a practical approach to find the total Hamiltonian
eigenvalues only through solving an algebra equation satisfied by
our kernal function, which is specially suitable to calculation in a
computer. In order to verify the validity of expression of
Hamiltonian eigenvalues and check the accuracy of calculation of
Hamiltonian eigenvalues using our approach, we study, as an example,
a quartic anharmonic oscillator in the normalization, which is
frequently used as a test-stone of new methods of calculating the
eigenvalues.

This article is organized in this way: in section~\ref{secttwo} we
give an overview on how to get our expression of the total
Hamiltonian eigenvalues in the general quantum systems. The detail
is put in four appendixes; in secton~\ref{sectthree} we present how
to obtain the total Hamiltonian eigenvalues by solving a given
algebraic equation; in section~\ref{sect4} we calculate the ground
state energy of the quartic anharmonic oscillator for different
coupling constants; in section~\ref{summary}, we make the summary.

\section{Derivation: Hamiltonian eigenvalues}\label{secttwo}

In the section, we would like to derive out an explicit expression
of Hamiltonian eigenvalues in a general quantum system. Actually,
our purpose is just to prove Wang's conjecture proposed in his
theoretical framework of quantum mechanics in the general quantum
systems\cite{wang1}.

In Wang's work, the complete series expression of time evolution
operator or transition amplitude in a solvable representation ($
H_0\ket{\Phi^\gamma}=E_\gamma \ket{\Phi^\gamma}$) is presented as
\beq \label{WangExpression} \bra{\Phi_\gamma}\e^{-\I H
t}\ket{\Phi_{\gamma^\prime}}= A^{\g \g'}=\sum_{l=0}^\infty
A_l^{\gamma\gamma^\prime} \eeq where \beqa \label{ADefinition}
A_0^{\gamma\gamma^\prime}&=& \e^{\I E_{\g}t}\delta_{\g\g'}\\
A_l^{\gamma\gamma^\prime}&=&\sum_{\g_1,\g_2,\ldots,\g_{l+1}}\sum_{i=1}^{l+1}\frac{\delta_{\g\g_1}\delta_{\g'\g_{l+1}}\e^{-\I
E_{\g_i}t}}{\displaystyle\prod_{j=1,j\neq
i}^{l+1}\left(E_{\g_i}-E_{\g_j}\right)} \prod_{k=1}^l
g^{\g_k\g_{k+1}} \eeqa It is clear that Wang's expression can be
thought of to be exact in the sense that this series involves
contributions from all order perturbations and has no any
approximation. In above expression, for simplicity, we have not
considered the degenerate cases. It is important that here we have
used the following subtle method of dividing Hamiltonian matrix in
the solvable $H_0$ representation \beq \{H\}=\{H_0\}+\{H_1\}=\left(
       \begin{array}{ccccc}
            E_0&    &       &      & \\
               &E_1 &       &      & \\
               &    &\ddots &      & \\
               &    &       &E_{\g}& \\
               &    &       &      &\ddots \\
       \end{array}
     \right)
     +\left(
       \begin{array}{cccccc}
         0        &g^{01} &g^{02} &\cdots     &g^{0 \g} &\cdots \\
         g^{10}   &0      &g^{12} &\cdots     &g^{1 \g} &\cdots\\
         g^{20}   &g^{21} &0      &\cdots     &\vdots   &\cdots\\
         \vdots   &\vdots &\vdots &\ddots     &g^{\g-1\g}&\cdots\\
         g^{\g 0} &g^{\g1}&\cdots &g^{\g\g-1} &0        &\cdots\\
         \vdots   &\vdots &\vdots &\vdots     &         &\ddots\\
       \end{array}
     \right)
\eeq Such an expression has separated the total Hamiltonian matrix
into diagonal part and off-diagonal part, rather than
nonperturbative part and perturbative part, which is marked with the
coupling constant in the normal perturbation theory. It is never
trivial in our derivations. Breaking the accustomed mentality in the
normal perturbation theory and using the more formalized expansion
form of time evolution operator are advantages of Wang's theory, and
they bring us successfully to arrive at our purpose.

Obviously, there are many apparent singular points in the expression
(\ref{ADefinition}) of $A_l^{\gamma\gamma^\prime}$ , but they are
fake in fact. In Wang's paper \cite{wang1} this problem has been
fixed by finding their limitations in terms of contraction and
anti-contraction of energy summation indexes. Here, we will theorize
Wang's method and further prove Wang's conjecture about the
eigenvalues of Hamiltonian.

A key trick using here is that we start from the partition function
and rewrite it by using Wang's expression (\ref{WangExpression})
\beq \sum_{\g}\e^{-\I\widetilde{E}{\g}t}=\sum_{\g}A^{\g\g} \eeq
where $ \widetilde{E}_{\g}$ are just the total Hamiltonian
eigenvalues that we would like to find. It is largely helpful for
removing the unexpected ``fake" singular points in Wang's framework,
which can be seen in Appendix \ref{AppendixA}. Another key skill is
that we use the contraction and anti-contraction of energy summation
indexes developed in \cite{wang1}. It makes all the apparent
singular points in the partition function expressed by Wang's
framework are neatly removed, which can be seen in the Appendix
\ref{AppendixB}. It is interesting that, based on the proof in
Appendix \ref{AppendixC} by particularly analyzing, skillfully
recombining the summations over energy indexes and perturbed order
indexes, we arrive at
\begin{equation}\label{PFExpression}
\sum_\gamma A^{\g \g}=\sum_{\g} \e^{-\I E{\g}t}
    \left\{1+(-\I t)
    \sum_{m=0}^\infty\frac{(-1)^{m}}{(m+1)!}\left[
    \left.\Dif{z}{m}
    \left(\e^{\I z t} R_{\g}^{m+1}(z) \right)
    \right]
    \right|_{z=0}\right\}
\end{equation}
where \beq R_\g(z)=\sum_{l=1}^\infty\sum_{\g_1\g_2\cdots\g_l\neq
\g}\frac{g^{\g\g_1}g^{\g_1\g_2}\cdots
g^{\g_l\g}}{(E_{\g}-E_{\g_1}-z)(E_{\g}-E_{\g_2}-z)\cdots(E_{\g}-E_{\g_l}-z)}
\eeq

$R_\g(z)$ is a kernal function that involves contributions from all
order perturbations, and plays a key role in our expression. It is
clear that $R_\g^{m+1}(z)$ and $\d^{m} R^{m+1}_\g(z)/dz^{m}|_{z=0}$
have the same order of magnitude of perturbed parameter $\lambda$,
but $R_\g^{m+1}(z)/R^m_\g(z)$ and $\left(\d^{m+1}
R_\g^{m+2}(z)/dz^{m+1}|_{z=0}\right)/\left(\d^{m}
R^{m+1}_\g(z)/dz^{m}|_{z=0}\right)$ is equal to
$\mathcal{O}(\lambda^2)$. This implies that the approximation
ability of this series is obviously improved and then it is more
suitable to cut upto some given the $m$th term since it decreases
more rapidly with $\lambda$.

However, more importantly and interestingly, we obtain \beq
\label{PFfinal}\sum_\g \e^{-\I \widetilde{E}_\g t}=\sum_\g
\exp\left\{-\I
\left({E}_\g+\sum_{m=0}^\infty\left.\frac{(-1)^{m}}{(m+1)!}\left[\Dif{z}{m}R_{\g}^{m+1}(z)\right]\right|_{z=0}\right)t\right\}
\eeq It is proved in the Appendix (\ref{AppendixD}) by expending the
partition function in Eq.(\ref{PFExpression}) into the time power
series and verifying the coefficient power relation. In fact, the
form of Eq.(\ref{PFfinal}) has its physics origin rather than the
mathematics arbitrariness, and it is valid in the general quantum
systems independent of the form of Hamiltonian. Therefore, we think
that the complete series expression of Hamiltonian eigenvalues in
the general quantum systems as below
\begin{equation}\label{EVexpression}
\widetilde{E}_\gamma=E_{\g}+\Delta E_\g=
E_\g+\sum_{m=0}^\infty\frac{(-1)^{m}}{(m+1)!}\left[\Dif{z}{m}R_{\g}^{m+1}(z)\right]\Bigg|_{z=0}
\end{equation}
Obviously, it is simply not a summation over the purturbed
parameter, but a series of power of the kernal function $R_\g(z)$ as
well as its derivative at $z=0$. It is completely different from the
normal perturbation theory in its thoughtway. In particular, when a
cut is introduced in a practical calculation, a higher $\lambda^2$
term than the last term of cut part is dropped, but not a higher
$\lambda$ term than the last term of cut part is dropped in the
normal perturbation theory.

It is easy to verify that our result is consistent with one in the
normal perturbation theory if one expands our expression
(\ref{EVexpression}) according to the order of perturbed parameter
for the known lower order forms in the textbook and refs. However,
our expression involves the contribution from all order perturbation
and has a neat form of general term. Rearranging summation in our
expression is helpful for theoretical derivation and practical
calculation since its completeness, orderliness and clearness. In
particular, it provides a physical reason how to chose part
contributions from higher order perturbations, which is able to
simplify the calculation and lead the result more precision. An
example has been presented in Ref.\cite{wang3}. In fact, our
rearrangement summation is reasonable because it reveals the
mathematical beauty and then physical nature. Moreover, its new
content, at least, to surprise us, will be obtained in the following
section.

Of course, our expression of the total Hamiltonian eigenvalues can
be thought of to be exact in the sense that this series involves the
contributions from all order perturbations and has no any
approximation and cut in form, as well as the general term is
obtained.

\section{Calculation: an algebra equation}\label{sectthree}

Although the complete series expression of total Hamiltonian
eigenvalues (\ref{EVexpression}) in the general quantum system
arrives at our theoretical aspiration and we believe that it is
interesting and important in the formulism of quantum mechanics, we
have to admit that this expression is probably too complicated to be
practical. It is inevitable and acceptable that the complete
expression of Hamiltonian eigenvalues (\ref{EVexpression}) is an
infinite series, because that a general quantum system has usually
no a compact solution without any approximation. It seems to be not
convenient in the calculation the $n$th order derivative of
$R_{\g}^{n+1}$, but because that $R_\g(z)$ is such a function with a
product form $1/(a_i-z)$, such a difficulty is not serious. However,
except for a reasonable rearrangement summation and a neat general
term to reasonably involve the higher order contributions, what is
more in our expression than one in the normal perturbation theory
for the calculation of total Hamiltonian eigenvalues. In this
section, our purpose is just designed to answer this problem.

Actually, we find that the difference between $H$'s and $H_0$'s
eigenvalues $\Delta E_\g$ is equal to the coefficient of $z$'s $0$th
power term in a Laurent series of $F_\g(z)$ as following
\beq\label{FGdifinition} F_\g(z)= z\ln\left(1+\frac{R_\g(z)}{
z}\right)\eeq It is easy to be proved. In fact, setting $u=
R_\g(z)/z$, we can rewrite $F_\gamma(z)=R_\g(z)\ln(1+u)/u$.
Obviously, the limit of $\ln(1+u)/u$ when $u\rightarrow 0$ is $1$.
Thus it means that $\ln(1+u)/u$ can be expanded as a Taylor's series
at $u=0$, that is \beq F_\g(z)=R_\g(z)\sum_{k=1}^\infty
\frac{(-1)^{k+1}}{k}
u^{k-1}=\sum_{k=0}^\infty\frac{(-1)^{k}}{k+1}\frac{R_\gamma^{k+1}}{z^k}
\eeq Again note that $R_\g(z)^{k+1}$ can be expanded as a Taylor's
series at $z=0$, we finish the proof of our above conclusion.

Now our task is to seek an explicit form of Laurent series of
$F_\g(z)$. We first set $\alpha_\g$ is a solution of following
equation \beq \label{EVformulaorig}R_\g(\alpha_\g)=-\alpha_\g\eeq
When $\alpha_\g$ is finite, from above equation it follows that
$R_\g(z)$ can be expended as\beq
R_\g(z)=R_\g(\alpha_\g)+\sum_{n=1}^\infty
\frac{1}{n!}\left[\Dif{\alpha_\g}{n}R_\g(\alpha_\g)\right]
(z-\alpha_\g)^n =- \alpha_\g+\overline{R}_\g(z)\eeq Then, we rewrite
\beqa\label{FgLSone} F_\g(z)&=& z\ln\left[\left(1-\frac{\alpha_\g}{
z}\right)+\frac{\overline{R}_\g(z)}{z}\right] =
z\ln\left[1-\frac{\alpha_\g}{z}\right]+
z\ln\left[1+\frac{\overline{R}_\g(z)}{(z-\alpha_\g)}\right] \eeqa
Since $\overline{R}_\g(z)/(z-\alpha_{\gamma})$ is canonical at
$z=\alpha_\g$ from Eq.({\ref{EVformulaorig}) for a finite
$\alpha_\g$, the second term in above equation does not involve
$z$'s 0th power part. Actually, this is a reason why to set
Eq.(\ref{EVformulaorig}). While the first term in Eq.(\ref{FgLSone})
can be expanded as \beqa
z\ln\left[1-\frac{\alpha_\g}{z}\right]&=&-\sum_{k=1}^\infty
\frac{z}{k}\left(\frac{\alpha_\g}{z}\right)^k\eeqa Therefore the
Laurent series of $F_\g(z)$ has its coefficient of z's 0th power to
be $-\alpha_\g$. It implies that we obtain \beq\label{EVformula}
\widetilde{E}_\g=E_\g+\Delta E_\g=E_\g-\alpha_\g\eeq This conclusion
is so interesting because it tells us that the total Hamiltonian
eigenvalues in the general quantum system can be calculated through
solving an algebra equation \beq R_\g(-\Delta E_\gamma)=\Delta
E_\g\eeq and then adding its solution $\Delta E_\g$ to the $H_0$'s
eigenvalues $E_\g$. More obviously, the total Hamiltonian
eigenvalues $\widetilde{E}_{\g}$ is a solution of following algebra
equation \beq \sum_{l=1}^\infty\sum_{\g_1\g_2\cdots\g_l\neq
\g}\frac{g^{\g\g_1}g^{\g_1\g_2}\cdots
g^{\g_l\g}}{(\widetilde{E}_\g-E_\g)(\widetilde{E}_{\g}-E_{\g_1})(\widetilde{E}_{\g}-E_{\g_2})\cdots(\widetilde{E}_{\g}-E_{\g_l})}
=1 \eeq This conclusion distinctly reveals the advantages of Wang's
formal framework and goes to an extent that we have ever not
researched using the known other theory. It must be emphasized that
the contributions from higher order even all order perturbation are
naturally involved in such an algebra equation. The difference
between our approach and usual one has been clearly seen here
because that we give up the accustomed calculation method order by
order in the normal perturbation theory.

\section{An example: quartic anharmonic oscillator}
\label{sect4}

In this section, we attempt to verify the validity of expression of
Hamiltonian eigenvalues and check the accuracy of calculation of
Hamiltonian eigenvalues using our method. As an example, we study a
quartic anharmonic oscillator in the normalization, which is
frequently used as a test-stone of new methods of calculating the
eigenvalues.

Harmonic oscillators and their anharmonic counter parts are
extremely important model systems in all branches of quantum physics
and particularly in quantum field theory. Even order anharmonic
oscillators defined by \beq H^{(m)}=p^2+x^2+\lambda x^{2m}, m=2,3,4
\cdots\eeq were studied in Bender and Wu's seminal work \cite{bender
and wu 1,bender and wu 2,bender and wu 3}, and then Simon made a
rigorous analysis of the mathematical property in \cite{simond}. The
perturbation expansions of the anharmonic oscillators diverge
strongly for the coefficients $c_m$ in Eq.(\ref{pee}) always grow
factorially. Therefore anharmonic oscillators are frequently used to
test new approximation techniques.

Let us begin to consider a quartic anharmonic oscillator. Its
Hamiltonian reads \beq H=-\Dif{x}{2}+x^2+\lambda x^4 \eeq Obviously,
the primary task is to write out our kernal function $R_{\g}$. But,
as it is defined, $R_{\g}$ is the summation of an infinite series.
Thus, a certain cut approximation should be made, that is, some
terms of this series need to be omitted. So, what strategy should be
adopted when trying to distinguish those of importance from trivial
terms? One may think of the tactics when making perturbation
approximation, we just pick those terms of low powers of the
coupling constant, and drop the higher powers. We are going to point
out that such a tactic won't be very proper here, for the two
reasons below. First, the usual perturbation expansion is a power
series of the coupling constant, but, our expression has no longer
been simply a power series of coupling constant. Second, even though
we admitted that the longer the term the smaller the value, the the
longer the term, the bigger the number of such terms. So, we propose
a new one, according to the fact that {\it the states nearby exert
larger impact than the states faraway} in our expression, we drop
the terms reflecting the effect of the states far away from the
initial one.

To put our idea into reality, we choose such some terms that the
summations over $\g_1,\g_2,\cdots,\g_l$ has a maximum value $n$ and
define them as a new series \beq {R}_{\g}^{\rm
c}(z,n)=\sum_{l=1}^{\infty}\sum_{\substack{
\g_1,\g_2\cdots\g_l\ne \g\\
\max\{\g_1,\g_2\cdots\g_l\}=n }} \frac{g^{\g\g_1}g^{\g_1\g_2}\cdots
g^{\g_l\g}}{(E_{\g}-E_{\g_1}-z)(E_{\g}-E_{\g_2}-
z)\cdots(E_{\g}-E_{\g_l}- z)}\eeq Thus, we can rewrite the kernal
function $R_{\g}$ as the summation of ${R}_{\g}^{\rm c}(z,n)$ over
$n$ as below \beq R_{\g}(z)=\sum_{n=1}^{\infty}{R}_{\g}^{\rm
c}(z,n)\eeq In the calculation of the ground state energy of quartic
anharmonc oscillator we make the cut approximation to $N$ \beq
R_{\g}(z)\approx\sum_{n=1}^{N}R_{\g}^{\rm c}(z,n)\eeq

To justify our approximation scheme, we've made some numerical
calculation as shown in Table.~\ref{tab:1}. And our result of the
ground state energy of quartic anharmonic oscillor for different
coupling constants is shown in Table.~\ref{tab:table2}

\begin{table}[h]
\caption{\label{tab:1}The value of $R_0^{\rm c}(0,n)$.}
\begin{ruledtabular}
\begin{tabular}{ccccccccc}
 $\lambda$ & $n=3$ & $n=5$    & $n=11$   & $n=21$   & $n=51$   & $n=101$   & $n=199$\\
 \hline
 0.1    & -9.1837e-003 &-4.5775e-004 &-1.4490e-007 &-5.0107e-012 &-8.1129e-024 &-2.8911e-041  &-4.3465e-068
\\
0.2 &-3.1034e-002 &-3.5160e-004 &-9.8794e-008 &-3.0709e-010
&-7.5163e-020 &-6.5572e-034  &-4.4807e-057 \\
1.0 &-3.4615e-001 &-2.1281e-002 &-8.6467e-004 &-1.5403e-006
&-6.5085e-013 &-4.5152e-022  &-3.4768e-040\\
2.0 &-8.1818e-001 &-1.2014e-001 &-1.5617e-003 &-8.4901e-007
&-2.3659e-011 &-4.2370e-018  &-4.4732e-032\\
10.0&-4.7872e+000 &-1.5066e+000 &-2.0541e-001 &-5.5894e-004
&-2.8380e-006 &-4.8053e-012  &-1.2683e-020\\
20.0&-9.7826e+000 &-3.4221e+000 &-1.5829e+000 &-7.2278e-001
&-1.3054e-003 &-3.1365e-008  &-1.8263e-014\\
100.0 &-4.9779e+001 &-1.8999e+001 &-3.0893e+001 &-1.2912e+001
&-4.2734e+002 &-3.6783e-002  &-5.2184e-006
\end{tabular}
\end{ruledtabular}
\end{table}

\begin{table}[h]
\caption{\label{tab:table2} The value of the ground state energy of
quartic anharmonic oscillator, using the the summation of the first
$n$ $R_0^{\rm c}(z,i)$ $(i=1,2\cdots,n)$ as the approximation of
$R_0(z)$.}
\begin{ruledtabular}
\begin{tabular}{cclccl}
 $\lambda$ &$n$   &$\widetilde{E}_0$                  &$\lambda$ &$n$   &$\widetilde{E}_0$
 \\\hline   0.1 &10 &1.06528570130099    &0.2 &10 &1.11829330436519\\
                &20 &1.06528550957781    &    &20 &1.11829265486895\\
                &30 &1.06528550957275    &    &30 &1.11829265444366\\
                &50 &1.06528550957275    &    &50 &1.11829265444348\\
                &100&1.06528550957275    &    &100&1.11829265444348\\
                &200&1.06528550957275    &    &200&1.11829265444348\\
                &known\footnotemark[1]  &1.0652855095437176888&    &known\footnotemark[1]    &1.1182926543670391534\\

   \hline   1.0 &10 &1.39337105560387    &2.0 &10 &1.61122760597946      \\
                &20 &1.39235392111137    &    &20 &1.60754799112121\\
                &30 &1.39235164865408    &    &30 &1.60754155853087\\
                &50 &1.39235164313030    &    &50 &1.60754130410410\\
                &100&1.39235164312960    &    &100&1.60754130407997\\
                &200&1.39235164312960    &    &200&1.60754130407997\\
                &known\footnotemark[1]  &1.3923516415302918557&    &known\footnotemark[1]    &1.6075413024685475387\\
   \hline   10.0&10 &2.47630097947871    &20.0&10 &3.18161125567721      \\
                &20 &2.45355539526673    &    &20 &3.02112722285399\\
                &30 &2.44923642985496    &    &30 &3.01172336279951\\
                &50 &2.44917490466071    &    &50 &3.00996284534114\\
                &100&2.44917407783815    &    &100&3.00994481629290\\
                &200&2.44917407782312    &    &200&3.00994481558327\\
                &known\footnotemark[1]  &2.4491740721183869183&    &known\footnotemark[1]    &3.0099448155577821983\\
   \hline 100.0 &10 &8.08464496277487    &&&\\
                &20 &5.14809927717057    &&&\\
                &30 &5.02007347355405    &&&\\
                &50 &5.00376751877937    &&&\\
                &100&4.99942534973870    &&&\\
                &200&4.99941754801155    &&&\\
                &known\footnotemark[1]  &4.9994175451375878293&&&\\
\end{tabular}
\end{ruledtabular}
\footnotetext[1]{Here, the value denoted by ``known" from
Ref.~\cite{known}.}
\end{table}
These results are good enough comparing with the known one in
\cite{known}.

\section{Summary}
\label{summary} %

In this article, we proposed a new approach calculating the
Hamiltonian eigenvalues in the general quantum systems. In
theoretical form, we obtain a complete series expression of total
Hamiltonian eigenvalues without any approximation and cut. In
practical calculation, we present an algebra equation satisfied by
the total Hamiltonian eigenvalues. These conclusions is based on
Wang's framework of quantum mechanics in the general quantum system.
Consequently we can say that Wang's formulism of quantum mechanics
in the general quantum systems is useful. In fact, the revised
Fermi's gold rule as well as its calculation \cite{wang3} also
accounts for Wang's theory has itself advantages.

Although only the non-degeneracy and discrete case is considered
here, but our derivation can be extended to the continuous and/or
degeneracy case.

It must be emphasized that an accustomed mentality that expands the
series according to the perturbed parameter in the normal
perturbation theory is given up, the contributions from all order
perturbation is involved via a kernal function, which plays a key
role in our theoretical expression and our calculation approach. The
advantages of our kernal function have been mentioned in the
previous sections. The calculation about this kernal function is an
important task. However, it only involves a product of matrices, and
then it can be efficiently calculated by using a computer. So in the
computability and usability, our approach is not weaker. In formal
beauty, please see our concrete expressions, no more words need to
say.

Perhaps, our expression is thought of a rearrangement of
perturbation series. However, such a rearrangement summation in form
is never trivial, it really shows the physics nature because it
directly comes from the law of quantum dynamics, but not a
transcendent input -- expansion according to the power of perturbed
parameter. In mathematics, the reasonable rearrangement of a
multiply series is often significant if a cut approximation is
needed. We think that the rearrangement summation in our expression
make us reasonably involve the contributions from part higher even
all order perturbations from our derivation. In fact, our expression
is more neat, more explicit and more deep than the known Kato's
\cite{Kato} and Bloch's \cite{Bloch}. This should be a reason why it
can lead to an interesting conclusion -- the total Hamiltonian
eigenvalues can be calculated by solving an algebra equation.

Our approach to calculate the total Hamiltonian eigenvalues also
gives up the accustomed way order by order, it makes the calculation
to be simplified, but still involves the contributions from
important part higher order perturbations when a cut approximation
is introduced. Actually, it is similar to choose the important parts
in higher order (equal to or more than the fourth order)
contributions based on some physics reasons. Specially, our equation
about the total Hamiltonian eigenvalues can more naturally and
conveniently involve the contributions from higher even all order
perturbations, and it is easy to be solved numerically in a
computer. Therefore, our approach is able to remarkably simplify the
calculation as well as advance the precision.

As an example, we applied our approach in the calculation of the
ground state energy of a quartic anharmonic oscillator. The highly
accurate results for the energies of the ground state energy of
quartic anharmonic oscillators of different coupling ranging from
0.1 to 100 are yielded. Something we should mention is that the
factorial divergence in the usual perturbation expansion of even
order anharmonic oscillators no longer appears in our expansion, and
our way of ``summing" expansion, which should be owned to the fact
that our expression is complete, is rather simple and neat in form
and also proved to be effective. On the basis of the results
presented in this article, one can expect that our new method should
also give good enough results in some other quantum mechanical
problems, and we'll proceed our work in the near future.

\begin{acknowledgments}
We are grateful all the collaborators of our quantum theory group in
our university. This work was supported by the National Natural
Science Foundation of China under Grant No. 10975125.
\end{acknowledgments}

\appendix

\section{}\label{AppendixA}

This appendix is focus on a derivation of the first expression of
partition function (\ref{PFExpression}) based Wang's formal
framework \cite{wang1}.

Note that the $H_1$ matrix is able to be taken as off-diagonal based
on Wang's proof \cite{wang1}, from Eq.(\ref{ADefinition}) it follows
\beq\label{A1}
A_1^{\gamma\gamma}=\sum_{\g_1,\g_2}\sum_{i=1}^{2}\frac{\e^{-\I
E_{\g_i}t}}{(E_{\g_1}-E_{\g_2})} \ g^{\g_1\g_2}
\delta_{\g\g_1}\delta_{\g\g_{2}}=0 \eeq It implies that the
contraction and anti-contraction of off-diagonal elements skill
using in \cite{wang1} is largely simplified when $H_1$ matrix is
taken as an off-diagonal form.

Since the existing the factor $\delta_{\g\g_1}\delta_{\g\g_{l+1}}$
in the $A_l^{\gamma\gamma}$ expression (\ref{ADefinition}), it leads
to that $E_{\g_1}=E_{\g_{l+1}}=E_{\g}$ after summation. It implies
that there is, at least, an obvious singular point among the
beginning and ending terms in the summation over $i$ from $1$ and
${l+1}$ for a fixed $l>1$. To remove it, we can combine this two
terms, and introduce an infinite small $\I\varepsilon$ to represent
the difference between $E_{\gamma_1}$ and $E_{\gamma_{l+1}}$, and
then replace $E_{\gamma_{1}}$ by $E_{\gamma}$, again finally set
$\varepsilon\rightarrow 0$. That is: \beqa \label{Alends}& &\left\{
    \frac{\e^{-\I E_{\g_1}t}}{\displaystyle\prod_{j=2}^{l+1}\left(E_{\g_i}-E_{\g_j}\right)}
    +\frac{\e^{-\I E_{\g_{l+1}}t}}{\displaystyle\prod_{j=1}^l\left(E_{\g_{l+1}}-E_{\g_j}\right)} \right\}
    \delta_{\g \g_1}\delta_{\g \g_{l+1}}\nonumber\\
& &\quad = \lim_{\varepsilon\to 0}\e^{-\I E_{\g}t}
    \left\{
    \frac{1}{\left[\displaystyle \prod_{i=2}^l(E_{\g}-E_{\g_i})\right](\I\varepsilon)}
    +\frac{\e^{-\varepsilon t}}{(-\I\varepsilon)
    \left[\displaystyle \prod_{i=2}^l(E_{\g}-E_{\g_i}-\I\varepsilon)\right]}\right\}
    \delta_{\g \g_1}\delta_{\g \g_{l+1}}\nonumber\\
& &\quad =-\e^{-\I E_{\g}t}
    \dif{z}\left.
    \left[\e^{\I z
    t}\prod_{i=2}^{l}\frac{1}{(E_{\g}-E_{\g_i}-z)}\right]\right|_{z=0}
    \delta_{\g \g_1}\delta_{\g \g_{l+1}}.
\eeqa The last step has directly used the definition of derivative.

Furthermore, let us first consider how to deal with the terms in the
summation over $i$ from $i=2$ to $i=l$ of Eq.(\ref{ADefinition}). By
summing the index $\gamma_1$ and $\gamma_{l+1}$ we have \beqa
\label{mterm}& &\sum_{\gamma,\gamma_1,\cdots,\gamma_{l+1}}
\sum_{i=2}^l\frac{\e^{-\I
E_{\gamma_i}t}}{\displaystyle\prod_{j=1,j\neq
i}^{l+1}\left(E_{\g_i}-E_{\g_j}\right)}\delta_{\gamma\gamma_1}\delta_{\gamma_{l+1}\gamma}\prod_{i=1}^l
g^{\gamma_i\gamma_{i+1}}=\sum_{\gamma,\gamma_2}\frac{\e^{-\I
E_{\gamma_2}t}}{(E_{\gamma_2}-E_{\gamma})^2} g^{\gamma\gamma_{2}}g^{\gamma_2\gamma}\delta_{l2}\nonumber\\
& & + \sum_{\gamma,\gamma_2,\gamma_{3}}\left[\frac{\e^{-\I
E_{\gamma_2}t}}{(E_{\gamma_2}-E_{\gamma})^2
(E_{\gamma_2}-E_{\gamma_3})}+\frac{\e^{-\I
E_{\gamma_3}t}}{(E_{\gamma_3}-E_{\gamma})^2
(E_{\gamma_3}-E_{\gamma_2})}\right]g^{\gamma\gamma_{2}}g^{\gamma_2\gamma_3}g^{\gamma_3\gamma}\delta_{l3}\nonumber\\
& & + \theta(l-4)\sum_{\gamma,\gamma_2,\cdots,\gamma_{l}}
\sum_{i=2}^{l}\left\{ \frac{\e^{-\I
E_{\gamma_i}t}}{(E_{\gamma_i}-E_{\gamma})^2
\left[\displaystyle\prod_{j=2}^{i-1}(E_{\gamma_i}-E_{\gamma_j})\right]\displaystyle\prod_{k=i+1}^{l}(E_{\gamma_i}-E_{\gamma_k})}\right\}
g^{\gamma\gamma_{2}}\left[\prod_{i=2}^{l-1}
g^{\gamma_i\gamma_{i+1}}\right]g^{\gamma_l\gamma}\eeqa where
$\theta(x)=1$ if $x\geq 0$, otherwise $\theta(x)=0$.

The simplest case is that $l=2$. Only there is the first term in
above equation (\ref{mterm}). Interchanging the dummy index
$\gamma\leftrightarrow\gamma_2$, we have \beqa \label{A2part}
\sum_{\gamma,\gamma_2}\frac{\e^{-\I
E_{\gamma_2}t}}{(E_{\gamma_2}-E_{\gamma})^2}
g^{\gamma\gamma_{2}}g^{\gamma_2\gamma}
&=&\sum_{\gamma,\gamma_2}\e^{-\I E_{\gamma}t} \left.\left[\e^{\I z
t}\dif{z}\left(\frac{1}{(E_{\gamma_2}-E_{\gamma}-
z)}\right)\right]\right|_{z=0}
g^{\gamma\gamma_{2}}g^{\gamma_2\gamma}\eeqa Similarly, for $l=3$, we
set $\{\gamma_2,\gamma_3,\gamma\}\rightarrow
\{\gamma,\gamma_2,\gamma_3\}$ for $i=2$, and
$\{\gamma_3,\gamma,\gamma_2\}\rightarrow
\{\gamma,\gamma_2,\gamma_3\}$ for $i=3$. Thus \beqa\label{A3part} &
&\sum_{\gamma,\gamma_1,\cdots,\gamma_{4}} \sum_{i=2}^3 \frac{\e^{-\I
E_{\gamma_i}t}}{\displaystyle\prod_{j=1,j\neq
i}^{4}\left(E_{\g_i}-E_{\g_j}\right)}\delta_{\gamma\gamma_1}\delta_{\gamma_{4}\gamma}\prod_{i=1}^3
g^{\gamma_i\gamma_{i+1}} \nonumber\\
& & = \sum_{\gamma,\gamma_2,\gamma_{3}}\left[\frac{\e^{-\I
E_{\gamma_2}t}}{(E_{\gamma_2}-E_{\gamma})^2
(E_{\gamma_2}-E_{\gamma_3})}+\frac{\e^{-\I
E_{\gamma_3}t}}{(E_{\gamma_3}-E_{\gamma})^2
(E_{\gamma_3}-E_{\gamma_2})}\right]
g^{\gamma\gamma_{2}}g^{\gamma_2\gamma_{3}}g^{\gamma_3\gamma}\nonumber\\
& & = \sum_{\gamma,\gamma_2,\gamma_{3}}\e^{-\I
E_{\gamma}t}\left[\frac{1}{(E_{\gamma}-E_{\gamma_2})^2
(E_{\gamma}-E_{\gamma_3})}+\frac{1}{(E_{\gamma}-E_{\gamma_2})
(E_{\gamma}-E_{\gamma_3})^2}\right]
g^{\gamma\gamma_{2}}g^{\gamma_2\gamma_{3}}g^{\gamma_3\gamma}\nonumber\\
& &= \sum_{\gamma,\gamma_2,\gamma_{3}}\e^{-\I
E_{\gamma}t}\left.\left\{\e^{\I z
t}\dif{z}\left[\frac{1}{(E_{\gamma}-E_{\gamma_2}-z)
(E_{\gamma}-E_{\gamma_3}-z)}\right]\right\}\right|_{z=0}
g^{\gamma\gamma_{2}}g^{\gamma_2\gamma_{3}}g^{\gamma_3\gamma}\eeqa
The skill of dummy index transformations can be continuously used
when $l\geq 4$. For $i=2$ and to $i=l$, they are respectively
$\{\gamma_2,\gamma_{3},\cdots,\gamma_{l},\gamma\}\rightarrow\{\gamma,\gamma_{2},\cdots,\gamma_{l-1},
\gamma_{l},\}$ and
$\{\gamma_{l},\gamma,\gamma_{2},\cdots,\gamma_{l-1}
\}\rightarrow\{\gamma,\gamma_{2},\cdots,\gamma_{l}\}$. For the other
$i$ ($l-1\geq i\geq 3$), our dummy index transformations are taken
as
$\{\gamma_i,\gamma_{i+1},\cdots,\gamma_{l},\gamma,\gamma_{2},\cdots,\gamma_{i-1}\}
\rightarrow\{\gamma,\gamma_{2},\cdots,\gamma_{l-i+1},\gamma_{l-i+2},\gamma_{l-i+3},\cdots,\gamma_{l}\}$.
It is easy to prove that, under above index transformations,
$g^{\gamma\gamma_{2}}g^{\gamma_2\gamma_3}\cdots g^{\gamma_l\gamma}$
is invariant in form. Obviously, $i=2$ and $i=l$ terms are
transformed as following form \beqa \frac{\e^{-\I
E_{\gamma_2}t}}{(E_{\gamma_2}-E_{\gamma})^2
\displaystyle\prod_{k=3}^{l}(E_{\gamma_2}-E_{\gamma_k})}&\rightarrow&
\frac{\e^{-\I E_{\gamma}t}}{\left[\displaystyle\prod_{k=2}^{l-1}(E_{\gamma}-E_{\gamma_k})\right](E_{\gamma}-E_{\gamma_l})^2}\\
\frac{\e^{-\I E_{\gamma_l}t}}{(E_{\gamma_l}-E_{\gamma})^2
\displaystyle\prod_{k=2}^{l-1}(E_{\gamma_l}-E_{\gamma_k})}&\rightarrow&
\frac{\e^{-\I E_{\gamma}t}}{(E_{\gamma}-E_{\gamma_2})^2
\displaystyle\prod_{k=3}^{l}(E_{\gamma}-E_{\gamma_k})}\eeqa While
the other $i$ from $3$ to $l-1$ ($l\geq 4$), our dummy index
transformations lead to \beqa \sum_{i=3}^{l-1}\frac{\e^{-\I
E_{\gamma_i}t}}{(E_{\gamma_i}-E_{\gamma})^2
\left[\displaystyle\prod_{j=2,j\neq
i}^{l}(E_{\gamma_i}-E_{\gamma_j})\right]}
\rightarrow\sum_{i=3}^{l-1}\frac{\e^{-\I E_{\gamma}t}}{
(E_{\gamma}-E_{\gamma_{i}})^2 \left[\displaystyle\prod_{j=2,j\neq
i}^{l}(E_{\gamma}-E_{\gamma_j})\right]}\eeqa Thus, when $l\geq 4$
\beqa & &\sum_{l=4}^{\infty}\sum_{\g,\g_1\cdots\g_{l+1}}
    \sum_{i=2}^{l}
    \frac{\e^{-\I E_{\g_i}t}}{\displaystyle\prod_{j=1,j\neq
i}^{l+1}\left(E_{\g_i}-E_{\g_j}\right)}\left[\prod_{k=1}^l
g^{\g_k\g_{k+1}}\right]\delta_{\g \g_1}\delta_{\g
    \g_{l+1}}\nonumber\\
    & &= \sum_{l=4}^{\infty}\sum_{\g,\g_2\g_3\cdots\g_{l}}\e^{-\I
E_{\gamma}t}\left\{\frac{1}{\left[\displaystyle\prod_{j=2}^{l-1}(E_{\gamma}-E_{\gamma_j})\right]
(E_{\gamma}-E_{\gamma_l})^2}+\sum_{i=3}^{l-1}\frac{1}{
(E_{\gamma}-E_{\gamma_{i}})^2
\left[\displaystyle\prod_{j=2,j\neq i}^{l}(E_{\gamma}-E_{\gamma_j})\right]}\right.\nonumber\\
& &\left.+ \frac{1}{(E_{\gamma}-E_{\gamma_2})^2
\displaystyle\prod_{k=3}^{l}(E_{\gamma}-E_{\gamma_k})}\right\}g^{\gamma\gamma_{2}}\left[\prod_{i=2}^{l-1}
g^{\gamma_i\gamma_{i+1}}\right]g^{\gamma_l\gamma} \nonumber\\
& &= \sum_{l=4}^{\infty}\sum_{\g,\g_2\g_3\cdots\g_{l}}\e^{-\I
E_{\gamma}t}\left.\left\{\e^{\I z
t}\dif{z}\left[\displaystyle\prod_{j=2}^{l}\frac{1}{(E_{\gamma}-E_{\gamma_j}-z)}\right]\right\}\right|_{z=0}
g^{\gamma\gamma_{2}}\left[\prod_{i=2}^{l-1}
g^{\gamma_i\gamma_{i+1}}\right]g^{\gamma_l\gamma}\label{Aeq4part}
\eeqa

Substituting
eqs.(\ref{A1},\ref{Alends},\ref{A2part},\ref{A3part},\ref{Aeq4part})
into the expression of partition function, we obtain
\beqa\label{PFsecond} \sum_{\gamma}A^{\gamma\gamma}&=&\sum_\gamma
\e^{-\I E_\gamma t}+ \sum_{l=2}^\infty
\sum_{\gamma,\gamma_2,\cdots,\gamma_l} \e^{-\I E_\gamma t}
g^{\gamma\gamma_{2}}
g^{\gamma_2\gamma_{3}}\cdots g^{\gamma_l\gamma} \nonumber\\
& &\times\left.\left\{\dif{z}
    \left[-\e^{\I z t}\prod_{j=2}^{l}\frac{1}{(E_{\g}-E_{\g_j}-
    z)}\right]+ \e^{\I z t}\dif{z}\left[\displaystyle\prod_{j=2}^{l}
\frac{1}{(E_{\gamma}-E_{\gamma_j}-z)}\right]\right\}\right|_{z=0}\nonumber\\
&=&\sum_\gamma \e^{-\I E_\gamma t}\left[1+ \sum_{l=1}^\infty
\sum_{\gamma_1,\cdots,\gamma_l} (-\I t)
g^{\gamma\gamma_{1}}g^{\gamma_1\gamma_2}\cdots
g^{\gamma_l\gamma}\prod_{j=1}^{l}
\frac{1}{(E_{\gamma}-E_{\gamma_j})}\right] \eeqa in which, we have
reset the summation indexes.

\section{}\label{AppendixB}

In this section, we would like to further remove all apparent
singular points in the expression of partition function.

It is clear that there are still the singular points in the
expression (\ref{PFsecond}) although they are fake. In fact, these
singular points as well as the relevant terms in form are similar to
those have been removed in Appendix \ref{AppendixA} except for the
delta function factors. Consequently, to further deal with these
unexpected singular points, we need to pick out the terms when
$E_{\gamma}=E_{\gamma_j}$. Its method is just like the things that
the ref.\cite{wang1} has ever done, that is, we start from the
identity in the sense of summation:\beq 1=
\delta_{\gamma\gamma_i}+\eta_{\gamma\gamma_j} \eeq and rewrite a
summation as \beq
\sum_{\gamma_i}f\left[x_\gamma,x_{\gamma_j}\right]=\sum_{\gamma_i}f\left[x_\gamma,x_{\gamma_i}\right]\delta_{\gamma\gamma_i}
+\sum_{\gamma_i}f\left[x_{\gamma},x_{\gamma_i}\right]\eta_{\gamma\gamma_i}
\eeq The first summation of right side of above equation is picked
out, in which, $x_{\gamma}=x_{\gamma_i}$. Extending this method to a
$l$-fold summation, we chose $m$ elements from a set
$\mathcal{L}=\{1,2,\cdots,l\}$ to form a subset $\{p^l\}_m$, so that
\beqa
1&=&\sum_{m=0}^{l}\sum_{\{p^l\}_m}\overline{\delta}_{\{p^l\}_m\gamma}\\
\overline{\delta}_{\{p^l\}_m\gamma}&=&\left[\prod_{i\in
\{p^l\}_m}\delta_{\gamma\gamma_{i}}\prod_{j\in
\{q^l\}_m}\eta_{\gamma\gamma_{j}}\right] \eeqa where
$\{q^l\}_m=\mathcal{L} -\{p^l\}_m$. Obviously \beqa
\{p^l\}_m&=&\{p^l_1,p^l_2,\cdots,p^l_m\in \mathcal{L}\; {\rm and}\;
p^l_i<p^l_j\; {\rm if}\; i<j\}\\
\{q^l\}_m&=&\{q^l_1,q^l_2,\cdots,q^l_{l-m}\in
(\mathcal{L}-\{p^l\}_m)\;{\rm and}\; q^l_i<q^l_j\;{\rm if}\; i<j\}
\eeqa In fact, if $p^l_i-p^l_j=1$, then
$g^{\gamma_i\gamma_j}\delta_{\gamma_i\gamma}\delta_{\gamma\gamma_j}=0$
since $g^{\g\g}=0$ has been taken here. This result leads to that
the number of subset $\{p^l\}_m$ element with contribution is not
larger than $\left[(l+1)/2\right]$. However, in form, we can keep
these vanishing terms.

In order to pick out the singular terms from the summation
$\displaystyle \sum_{\gamma_1,\cdots,\gamma_l}\e^{-\I E_\g
t}{\displaystyle \prod_{j=1}^{l} (E_{\gamma}-E_{\gamma_j})^{-1}}$ we
product it by $\overline{\delta}_{\{p^l\}_m\gamma}$ so that there
are, at least, $m$ obvious singular points within it since
$\overline{\delta}_{\{p^l\}_m\gamma}$ contains $m$ delta functions.
Obviously, $m=0$ case is not needed to considered since there is no
the singular point.

Similar to the skill used in Appendix \ref{AppendixA}, we rewrite it
as $(m+1)$ terms, in which, the first term does not involves the
delta function action, but each in the other $m$ terms respectively
absorbs the contribution of each delta function, that is \beqa &
&\left[\frac{\e^{-\I E_{\gamma} t}}{\displaystyle \prod_{j=1}^{l}
(E_{\gamma}-E_{\gamma_j})}\right]\overline{\delta}_{\{p^l\}_m\g}=\frac{1}{m+1}\left[\frac{\e^{-\I
E_{\gamma} t}}{\displaystyle \prod_{j=1}^{l}
(E_{\gamma}-E_{\gamma_j})}+\frac{\e^{-\I E_{\gamma_{p_1}}
t}}{(E_{\gamma_{p_1}}-E_{\gamma})\displaystyle \prod_{j=1,j\neq
p_1}^{l} (E_{\gamma_{p_1}}-E_{\gamma_j})}\right. \nonumber\\
& & \left.+\frac{\e^{-\I E_{\gamma_{p_2}}
t}}{(E_{\gamma_{p_2}}-E_{\gamma})\displaystyle \prod_{j=1,j\neq
p_2}^{l} (E_{\gamma_{p_2}}-E_{\gamma_j})}+\cdots +\frac{\e^{-\I
E_{\gamma_{p_m}} t}}{(E_{\gamma_{p_m}}-E_{\gamma})\displaystyle
\prod_{j=1,j\neq p_m}^{l}
(E_{\gamma_{p_m}}-E_{\gamma_j})}\right]\overline{\delta}_{\{p^l\}_m\gamma}
\eeqa In order to remove the fake singularity, we introduce $m$
infinite small numbers $\varepsilon_m$ $(m=1,2,\cdots,l)$ so that
$E_\gamma-E_{\gamma_{p_m}}=\I\varepsilon_m$. Using the fact that
$E_{\gamma_{p_i}}-E_{\gamma_{p_j}}=-\I(\varepsilon_i-\varepsilon_j)$
and noting the action within $\overline{\delta}_{\{p^l\}_m\g}$ , we
have \beqa & &\left[\frac{\e^{-\I E_{\gamma} t}}{\displaystyle
\prod_{j=1}^{l}
(E_{\gamma}-E_{\gamma_j})}\right]\overline{\delta}_{\{p^l\}_m\gamma}=\frac{\e^{-\I
E_{\gamma} t}}{m+1}
\lim_{\substack{\varepsilon_1\rightarrow 0 \\
\varepsilon_2\rightarrow 0
\\\cdots\\  \varepsilon_m\rightarrow 0}}
\left[\frac{1}{\I^m\displaystyle\prod_{i=1}^m
\left(\varepsilon_i\right)}\frac{1}{\displaystyle \prod_{j=1}^{l-m}
(E_{\gamma}-E_{\gamma_{q^l_j}})}\right.\nonumber\\
& &\left. -\sum_{k=1}^m
\frac{1}{\I^m\varepsilon_k\displaystyle\prod_{i=1,i\neq k}^{m}
\left(\varepsilon_{i}-\varepsilon_k\right)}\frac{\e^{-\varepsilon_k
t}}{\displaystyle \prod_{j=1}^{l-m}
(E_{\gamma}-E_{\gamma_{q^l_j}}-\I\varepsilon_k)}\right]\overline{\delta}_{\{p^l\}_m\gamma}\\
&=&\frac{(-\I)^m\e^{-\I E_{\gamma} t}}{m+1} \lim_{\varepsilon_1,\
\cdots, \varepsilon_m\rightarrow 0}\sum_{k=0}^m
B_k\overline{\delta}_{\{p^l\}_m\gamma}\label{BFirsteq}
 \eeqa where we have denoted the $q^l_j \in \{q^l\}_m$
$(j=1,2,\cdots,l-m)$, rewrite the summation or production over the
index belonging to $\{q^l\}_m$ (or $\not\in \{p^l\}_m$) as from
$q^l_1$ to $q^l_{l-m}$, and also define $B_k$ by \beqa
B_0&=&\frac{1}{\displaystyle\prod_{i=1}^m
\left(\varepsilon_i\right)}\frac{1}{\displaystyle
\prod_{j=1}^{l-m} (E_{\gamma}-E_{\gamma_{q^l_j}})}\\
B_k&=&\frac{1}{\varepsilon_k\displaystyle\prod_{i=1,i\neq k}^{m}
\left(\varepsilon_{i}-\varepsilon_k\right)}\frac{\e^{-\varepsilon_k
t}}{\displaystyle \prod_{j=1}^{l-m}
(E_{\gamma}-E_{\gamma_{q^l_j}}-\I\varepsilon_k)}\quad (k\geq 1)
\eeqa

Now let us prove the following equation \beq\label{BSecondeq}
\left[\frac{\e^{-\I E_{\gamma} t}}{\displaystyle \prod_{j=1}^{l}
(E_{\gamma}-E_{\gamma_j})}\right]\overline{\delta}_{\{p^l\}_m\gamma}=\e^{-\I
E_{\g}t}
    \frac{(-1)^{m}}{(m+1)!}
    \left.\Dif{z}{m}
    \left[\prod_{j=1}^{l-m} \frac{\e^{\I zt}}{(E_{\g}-E_{\g_{q_j^l}}-z)}
    \right]\right|_{z=0}\overline{\delta}_{\{p^l\}_m\g}. \eeq

Firstly, we calculate the limitation $\varepsilon_1\rightarrow 0$.
Only $\varepsilon_1$ is an obvious singular point in $B_0$ and
$B_1$, that is \beqa\label{Bstep1one} \lim_{\varepsilon_1\rightarrow
0} \left(\sum_{k=0}^m
B_k\right)\overline{\delta}_{\{p^l\}_m\gamma}&=&
\lim_{\varepsilon_1\rightarrow 0}
\left[\frac{1}{\varepsilon_1\displaystyle\prod_{i=2}^m
\left(\varepsilon_i\right)}\frac{1}{\displaystyle \prod_{j=1}^{l-m}
(E_{\gamma}-E_{\gamma_{q^l_j}})}-\frac{1}{\varepsilon_1\displaystyle\prod_{i=2}^{m}
\left(\varepsilon_{i}-\varepsilon_1\right)}\frac{\e^{-\varepsilon_1
t}}{\displaystyle \prod_{j=1}^{l-m}
(E_{\gamma}-E_{\gamma_{q^l_j}}-\I\varepsilon_1)}\right]\overline{\delta}_{\{p^l\}_m\gamma}\nonumber\\
& &+\left[\sum_{k=2}^m
\frac{1}{\varepsilon_k^2\displaystyle\prod_{i=2,i\neq k}^{m}
\left(\varepsilon_i-\varepsilon_k\right)}\frac{\e^{-\varepsilon_k
t}}{\displaystyle \prod_{j=1}^{l-m}
(E_{\gamma}-E_{\gamma_{q^l_j}}-\I\varepsilon_k)}\right]\overline{\delta}_{\{p^l\}_m\gamma}
\eeqa The last summation term appears only when $m\geq 2$. Actually,
by setting \beq\label{myfunction}
f_{lm}^k(x)=\frac{1}{\displaystyle\prod_{i=k}^{m}
\left(\varepsilon_{i}-x\right)}\frac{\e^{-x t}}{\displaystyle
\prod_{j=1}^{l-m} (E_{\gamma}-E_{\gamma_{q^l_j}}-\I x)}\eeq next
doing its Taylor expansion \beqa\label{fte}
f_{lm}^k(x)&=&f_{lm}(0)+\left.\left[\dif{x}f_{lm}^k(x)\right]\right|_{x=0}x+
\frac{1}{2!}\left.\left[\Dif{x}{2}f_{lm}^k(x)\right]\right|_{x=0}x^2+\mathcal{O}(x^3)\nonumber\\
&=&\frac{1}{\displaystyle\prod_{i=k}^{m}
\varepsilon_{i}}\frac{1}{\displaystyle \prod_{j=1}^{l-m}
(E_{\gamma}-E_{\gamma_{q^l_j}})}+\left.\dif{x}\left[\frac{1}{\displaystyle\prod_{i=k}^{m}
\left(\varepsilon_{i}-x\right)}\frac{1}{\displaystyle
\prod_{j=1}^{l-m}
(E_{\gamma}-E_{\gamma_{q^l_j}}-\I x)}\right]\right|_{x=0}x\nonumber\\
&
&+\left.\frac{1}{2!}\Dif{x}{2}\left[\frac{1}{\displaystyle\prod_{i=k}^{m}
\left(\varepsilon_{i}-x\right)}\frac{\e^{-x t}}{\displaystyle
\prod_{j=1}^{l-m} (E_{\gamma}-E_{\gamma_{q^l_j}}-\I
x)}\right]\right|_{x=0}x^2+\mathcal{O}(x^3) \eeqa and then
substituting $k=2$ result into Eq.(\ref{Bstep1one}), we arrive at

\beqa\label{Bstep1two} \lim_{\varepsilon_1\rightarrow 0}
\left(\sum_{k=0}^m B_k\right)\overline{\delta}_{\{p^l\}_m\gamma}&=&
(-1)\dif{\varepsilon_1}\left.\left[\frac{1}{\displaystyle\prod_{i=2}^{m}
\left(\varepsilon_{i}-\varepsilon_1\right)}\frac{\e^{-\varepsilon_1
t}}{\displaystyle \prod_{j=1}^{l-m}
(E_{\gamma}-E_{\gamma_{q^l_j}}-\I\varepsilon_1)}\right]\right|_{\varepsilon_1=0}\overline{\delta}_{\{p^l\}_m\gamma}\nonumber\\
& &+\left[\sum_{k=2}^m
\frac{1}{\varepsilon_k^2\displaystyle\prod_{i=2,i\neq k}^{m}
\left(\varepsilon_i-\varepsilon_k\right)}\frac{\e^{-\varepsilon_k
t}}{\displaystyle \prod_{j=1}^{l-m}
(E_{\gamma}-E_{\gamma_{q^l_j}}-\I\varepsilon_k)}\right]\overline{\delta}_{\{p^l\}_m\gamma}
\eeqa From the definition of derivative of a function, this result
is very obvious.

Secondly, let us find the limitation $\varepsilon_2\rightarrow 0$.
It is easy to see \beqa\label{Bstep2one}
\lim_{\varepsilon_1,\varepsilon_2\rightarrow 0} \left(\sum_{k=0}^m
B_k\right)\overline{\delta}_{\{p^l\}_m\gamma}&=&\lim_{\varepsilon_2\rightarrow
0}\left[
(-1)\dif{\varepsilon_1}f_{lm}^2(\varepsilon_1)\big|_{\varepsilon_1=0}
+\frac{1}{\varepsilon_2^2}f_{lm}^3(\varepsilon_2)\right]\overline{\delta}_{\{p^l\}_m\gamma}\nonumber\\
& &-\left[\sum_{k=3}^m
\frac{1}{\varepsilon_k^3\displaystyle\prod_{i=3,i\neq k}^{m}
\left(\varepsilon_i-\varepsilon_k\right)}\frac{\e^{-\varepsilon_k
t}}{\displaystyle \prod_{j=1}^{l-m}
(E_{\gamma}-E_{\gamma_{q^l_j}}-\I\varepsilon_k)}\right]\overline{\delta}_{\{p^l\}_m\gamma}
\eeqa where $f_{lm}^k(x)$ is defined by Eq.(\ref{myfunction}). From
\beq
(-1)\dif{\varepsilon_1}\left.\left[f_{lm}^2(\varepsilon_1)\right]\right|_{\varepsilon_1=0}
=-\frac{1}{\varepsilon_2^2}f_{lm}^3(0)
-\frac{1}{\varepsilon_2}\left.\left[\dif{\varepsilon_1}f_{lm}^3(\varepsilon_1)\right]\right|_{\varepsilon_1=0}
\eeq and Eq.(\ref{fte}) but taking $k=3$, it follows that \beqa
\lim_{\varepsilon_1,\varepsilon_2\rightarrow 0} \left(\sum_{k=0}^m
B_k\right)\overline{\delta}_{\{p^l\}_m\gamma}&=&\frac{(-1)^2}{2!}\Dif{\varepsilon_2}{2}\left.\left[\frac{1}{\displaystyle\prod_{i=3}^{m}
\left(\varepsilon_{i}-\varepsilon_2\right)}\frac{\e^{-\varepsilon_2
t}}{\displaystyle \prod_{j=1}^{l-m}
(E_{\gamma}-E_{\gamma_{q^l_j}}-\I\varepsilon_2)}\right]\right|_{\varepsilon_2=0}\overline{\delta}_{\{p^l\}_m\gamma}\nonumber\\
& &-\left[\sum_{k=3}^m
\frac{1}{\varepsilon_k^3\displaystyle\prod_{i=3,i\neq k}^{m}
\left(\varepsilon_i-\varepsilon_k\right)}\frac{\e^{-\varepsilon_k
t}}{\displaystyle \prod_{j=1}^{l-m}
(E_{\gamma}-E_{\gamma_{q^l_j}}-\I\varepsilon_k)}\right]\overline{\delta}_{\{p^l\}_m\gamma}
\eeqa In particular, \beqa
\lim_{\varepsilon_1,\varepsilon_2\rightarrow 0} \left(\sum_{k=0}^2
B_k\right)\overline{\delta}_{\{p^l\}_2\gamma}&=&
\frac{(-1)^2}{2!}\Dif{\varepsilon_2}{2}\left.\left[\frac{\e^{-\varepsilon_2
t}}{\displaystyle \prod_{j=1}^{l-2}
(E_{\gamma}-E_{\gamma_{q^l_j}}-\I\varepsilon_2)}\right]\right|_{\varepsilon_2=0}\overline{\delta}_{\{p^l\}_2\gamma}\eeqa

Analyzing the above derivation, we have seen the fact that the
singular points are able be removed by finding limitations step by
step. Every limitation calculation increases one order of derivative
in the first term and decreases a term in the last summation. When
all limitations are calculated, this expression becomes a pure $m$
order derivative. Without loss of generality, for $n\leq (m-2)$ we
assume that \beqa\label{Bstepn}
\lim_{\substack{\varepsilon_1\rightarrow 0
\\\varepsilon_2\rightarrow 0\\   \cdots\\
\varepsilon_{n}\rightarrow 0}} \left(\sum_{k=0}^m
B_k\right)\overline{\delta}_{\{p^l\}_m\gamma}&=&\frac{(-1)^n}{n!}
\Dif{\varepsilon_n}{n}\left.\left[f_{lm}^{n+1}(\varepsilon_n)\right]\right|_{\varepsilon_n=0}\overline{\delta}_{\{p^l\}_m\gamma}\nonumber\\
& &+\left[\sum_{k={n+1}}^m
\frac{1}{(-\varepsilon_k)^{n+1}\displaystyle\prod_{i=n+1,i\neq
k}^{m}
\left(\varepsilon_i-\varepsilon_k\right)}\frac{\e^{-\varepsilon_k
t}}{\displaystyle \prod_{j=1}^{l-m}
(E_{\gamma}-E_{\gamma_{q^l_j}}-\I\varepsilon_k)}\right]\overline{\delta}_{\{p^l\}_m\gamma}
\eeqa we can have \beqa\label{Bstepnadd1}
\lim_{\substack{\varepsilon_1\rightarrow 0
\\\varepsilon_2\rightarrow 0\\   \cdots\\
\varepsilon_{n+1}\rightarrow 0}} \left(\sum_{k=0}^m
B_k\right)\overline{\delta}_{\{p^l\}_m\gamma}&=&\lim_{\varepsilon_{n+1}\rightarrow
0}\left\{\frac{(-1)^n}{n!}
\Dif{\varepsilon_n}{n}\left.\left[f_{lm}^{n+1}(\varepsilon_n)\right]\right|_{\varepsilon_n=0}
+\frac{(-1)^{n+1}}{\varepsilon_{n+1}^{n+1}}f_{lm}^{n+2}(\varepsilon_{n+1})\right\}\overline{\delta}_{\{p^l\}_m\gamma}\nonumber\\
& & +\left[\sum_{k={n+2}}^m
\frac{1}{(-\varepsilon_k)^{n+2}\displaystyle\prod_{i=n+2,i\neq
k}^{m}
\left(\varepsilon_i-\varepsilon_k\right)}\frac{\e^{-\varepsilon_k
t}}{\displaystyle \prod_{j=1}^{l-m}
(E_{\gamma}-E_{\gamma_{q^l_j}}-\I\varepsilon_k)}\right]\overline{\delta}_{\{p^l\}_m\gamma}
\eeqa Using the fact that \beqa
\Dif{\varepsilon_n}{n}\left.\left[f_{lm}^{n+1}(\varepsilon_n)\right]\right|_{\varepsilon_n=0}&=&
\Dif{\varepsilon_n}{n}\left.\left[\frac{1}{\varepsilon_{n+1}-\varepsilon_n}f_{lm}^{n+2}(\varepsilon_n)\right]\right|_{\varepsilon_n=0}
=\sum_{j=0}^n\frac{n!}{j!}
\frac{1}{\varepsilon_{n+1}^{n+1-j}}\left.\Dif{\varepsilon_{n}}{j}f_{lm}^{n+2}(\varepsilon_n)\right|_{\varepsilon_n=0}\\
f_{lm}^{n+2}(\varepsilon_{n+1})&=&\sum_{j=0}^\infty
\frac{1}{j!}\left.\Dif{\varepsilon_{n}}{j}f_{lm}^{n+2}(\varepsilon_{n})\right|_{\varepsilon_n=0}\varepsilon_{n+1}^j
 \eeqa
we have
 \beqa\label{Bstepnadd2}
\lim_{\substack{\varepsilon_1\rightarrow 0
\\\varepsilon_2\rightarrow 0\\   \cdots\\
\varepsilon_{n+1}\rightarrow 0}} \left(\sum_{k=0}^m
B_k\right)\overline{\delta}_{\{p^l\}_m\gamma}&=&\frac{(-1)^{n+1}}{(n+1)!}\left.\left[\Dif{\varepsilon_{n+1}}{n+1}
f_{lm}^{n+2}(\varepsilon_{n+1})\right]\right|_{\varepsilon_{n+1}=0}\overline{\delta}_{\{p^l\}_m\gamma}\nonumber\\
& & +\left[\sum_{k={n+2}}^m
\frac{1}{(-\varepsilon_k)^{n+2}\displaystyle\prod_{i=n+2,i\neq
k}^{m}
\left(\varepsilon_i-\varepsilon_k\right)}\frac{\e^{-\varepsilon_k
t}}{\displaystyle \prod_{j=1}^{l-m}
(E_{\gamma}-E_{\gamma_{q^l_j}}-\I\varepsilon_k)}\right]\overline{\delta}_{\{p^l\}_m\gamma}
\eeqa Finally, set $n=m-1$ and then use the same method we can
finish the proof of Eq.(\ref{BSecondeq}). Obviously \beqa
\lim_{\substack{\varepsilon_1\rightarrow 0
\\\varepsilon_2\rightarrow 0\\   \cdots\\
\varepsilon_{m}\rightarrow 0}} \left(\sum_{k=0}^m
B_k\right)\overline{\delta}_{\{p^l\}_m\gamma}&=&\lim_{\varepsilon_m\rightarrow
0}\left[\frac{(-1)^{m-1}}{(m-1)!}\left.\left[\Dif{\varepsilon_{m-1}}{m-1}
f_{lm}^{m}(\varepsilon_{m-1})\right]\right|_{\varepsilon_{m-1}=0}+\frac{(-1)^m}{\varepsilon_m^{m}}\frac{\e^{-\varepsilon_m
t}}{\displaystyle \prod_{j=1}^{l-m}
(E_{\gamma}-E_{\gamma_{q^l_j}}-\I\varepsilon_m)}\right]\overline{\delta}_{\{p^l\}_m\gamma}\nonumber\\
&=& \frac{(-1)^{m}}{m!}\left.\Dif{\varepsilon}{m}
\left[\frac{\e^{-\varepsilon t}}{\displaystyle \prod_{j=1}^{l-m}
(E_{\gamma}-E_{\gamma_{q^l_j}}-\I\varepsilon)}\right]\right|_{\varepsilon=0}\overline{\delta}_{\{p^l\}_m\gamma}\eeqa
Substituting
it into Eq.(\ref{BFirsteq}) and setting $\I\varepsilon=z$ follow the
conclusion Eq.(\ref{BSecondeq})

When the set $\{q^l\}_l$ is not an empty set, the product or
summation over this set is well-defined. Usually, in the formal
expressions, the product over an empty is thought of $1$ and the
summation over an empty is thought of $0$. At the above sense, $m$
still can take $l$. But $g^{\g\g}=0$ has been taken, thus \beq
g^{\g\g_1}g^{\g_1\g_2}\cdots
g^{\g_l\g}\overline{\delta}_{\{p^l\}_{l}\g}=0\eeq

Finally, we arrive at

\beqa \label{ABPFfirst} \sum A^{\g \g}&=&\sum_{\g} \e^{-\I E{\g}t}
+\sum_{l=1}^{\infty}\sum_{m=0}^{l-1}\sum_{\{p^l\}_m}\sum_{\g\g_1\g_2\cdots\g_l}
    (-it)\e^{-\I E_{\g}t}
    \left.\frac{(-1)^{m}}{(m+1)!}
    \Dif{z}{m}
    \left[ \frac{\e^{\I z t}}{\displaystyle \prod_{j=1,j\not\in\{p^l\}_m}^{l} (E_{\g}-E_{\g_j}-z)}
    \right]\right|_{z=0}\nonumber\\
& & g^{\g\g_1}g^{\g_1\g_2}\cdots
g^{\g_l\g}\overline{\delta}_{\{p^l\}_m\g} \eeqa It must be
emphasized that all obvious singular points have been removed in
above form.

\section{}\label{AppendixC}

Now, our task is to continue to derive out the final expression of
partition function that is written as a series of power of a kernal
function as well as its derivative.

It is clear that $\{p^l\}_m$ is a subset of
$\mathcal{L}=\{1,2,\cdots,l\}$, and has, at most, $l-1$ elements.
Its every element can be taken as an end point so that $\mathcal{L}$
is divided into $m+1$ subsets
$\mathcal{L}_n=\{p^l_{n-1}+1,p^l_{n-1}+2,\cdots,p^l_n\}$
($n=1,2,\cdots,m+1$) except for $p_m=l$ case. But this exceptional
case does not really appear since
$g^{\g_{p_m}\g}\overline{\delta}_{\{p^l\}_m\g}=g^{\g\g}\overline{\delta}_{\{p^l\}_m\g}=0$
by using the fact that $g^{\g\g}$ has been taken as off diagonal,
but in form, we simply set $\mathcal{L}_{m+1}$ to be an empty set.

Obviously, the cardinal numbers of $\mathcal{L}_n$ are respectively
$l_n=p^l_n-p^l_{n-1}$, in which $p^l_0=0$ and $p^l_{m+1}=l$ are
defined. For unify the denotation we need to define the subsets that
do not contain the $p^l_i$, they are
$\mathcal{L}^\prime_n=\mathcal{L}_n-p_n^l$ and in the exceptional
case when $p_m=l$, $\mathcal{L}^\prime_{m+1}= \varnothing$. Their
cardinal numbers are respectively $l^\prime_n=l_n-1$. But in the
exceptional case when $p_m=l$, we still set $l^\prime_{m+1}=0$.

After summing all delta functions within
$\overline{\delta}_{\gamma_{\{p^l\}_m\gamma}}$ and grouping the
relevant terms, we have \beqa &
&\sum_{\g_1,\cdots,\g_l}\overline{\delta}_{\{p^l\}_m\gamma}
g^{\g\g_1}g^{\g_1\g_2}\cdots g^{\g_l\g} \prod_{i=1,i\not\in
\{p^l\}_m}^l\frac{1}{(E_\gamma-E_{\gamma_i}-z)}\nonumber\\
& &
=\prod_{n=1}^{m+1}\left[\sum_{\g_{p^l_n+1},\cdots,\g_{p^l_n-1}}g^{\g\g_{p^l_{n-1}+1}}g^{\g_{p^l_{n-1}+1}\g_{p^l_{n-1}+2}}\cdots
g^{\g_{p^l_n-1}\g}
\prod_{i=p^l_{n-1}+1}^{p_n^1-1}\frac{\eta_{\gamma\gamma_i}}{(E_\gamma-E_{\gamma_i}-z)}\right]\\
\eeqa It must be emphasized that $l^\prime_n=0$ will lead to appear
the factor
$g^{\g_{p^l_i}\g_{p^l_{i+1}}}\delta_{\g_{p^l_i}\g}\delta_{\g_{p^l_{i+1}}\g}
=g^{\g\g}\delta_{\g_{p^l_i}\g}\delta_{\g_{p^l_{i+1}}\g}=0$, and no
any
$\displaystyle\frac{\eta_{\gamma\gamma_i}}{(E_\gamma-E_{\gamma_i}-z)}$
is grouped into the square bracket. Note that the dummy indexes can
be changed according to the given rules, we define \beq
R_\gamma^{(l^\prime_n)}(z)=\sum_{\g_1,\cdots,\g_{l^\prime_n}}g^{\g\g_1}g^{\g_1\g_2}\cdots
g^{\g_{l^\prime_n}\g}
\prod_{i=1}^{l^\prime_n}\frac{\eta_{\g_i\g}}{(E_\gamma-E_{\gamma_i}-z)}
\eeq and then \beq
\sum_{\g_1,\cdots,\g_l}\overline{\delta}_{\{p^l\}_m\gamma}
g^{\g\g_1}g^{\g_1\g_2}\cdots g^{\g_l\g} \prod_{i=1,i\not\in
\{p^l\}_m}^l\frac{1}{(E_\gamma-E_{\gamma_i}-z)}=\prod_{n=1}^{m+1}
R^{(l^\prime_n)}_\gamma(z) \eeq

Again substituting it into Eq.(\ref{ABPFfirst}), we arrive at \beqa
\sum A^{\g \g}&=&\sum_{\g} \e^{-\I E{\g}t}
+\sum_\gamma\sum_{l=1}^{\infty}\sum_{m=0}^{l-1}\sum_{\{p^l\}_m}
    (-\I t)\e^{-\I E_{\g}t}
    \left.\frac{(-1)^{m}}{(m+1)!}
    \Dif{z}{m}
    \left[\prod_{n=1}^{m+1}
R^{(l^\prime_n)}_\gamma(z)\e^{\I z t}
    \right]\right|_{z=0} \eeqa
The summations over all set $\{p^l\}_m$ can be changed into the
summations over $l^\prime_1,l^\prime_2,\cdots,l^\prime_m$, but there
is a limitation that $\sum_{n=1}^m l_n\leq l$. Interchanging the
summations over $l$ and $m$, then $l$ begins at $m+1$, setting
$l^\prime_{m+1}=l-m$, and noting $l^\prime_{m+1}$ can be taken up to
infinity, every $l^\prime_n$ ($n=1,2,\cdots,m)$ also can be taken up
to infinity. Therefore \beqa\label{ABPFthird} \sum A^{\g
\g}&=&\sum_{\g} \e^{-\I E{\g}t}
+\sum_{\gamma}\sum_{m=0}^{\infty}\sum_{l^\prime_1,\cdots,l^\prime_{m+1}=1}^\infty
    (-\I t)\e^{-\I E_{\g}t}
    \left.\frac{(-1)^{m}}{(m+1)!}
    \Dif{z}{m}
    \left[\prod_{n=1}^{m+1}
R^{(l^\prime_n)}_\gamma(z)
    \e^{\I z t}\right]\right|_{z=0}\nonumber\\
&=&\sum_{\g} \e^{-\I E{\g}t}\left\{1+\sum_{m=0}^{\infty}
    (-\I t)
    \left.\frac{(-1)^{m}}{(m+1)!}
    \Dif{z}{m}\left[
R_\gamma^{m+1}(z)
    \e^{\I z t}\right]\right|_{z=0}\right\} \eeqa
where we define the function $R_{\g}(z)$ by \beq
R_\g(z)=\sum_{l=1}^\infty
R_\gamma^{(l)}(z)=\sum_{l=1}^\infty\sum_{\g_1\g_2\cdots\g_l\neq\g}\frac{g^{\g\g_1}g^{\g_1\g_2}\cdots
g^{\g_l\g}}{(E_{\g}-E_{\g_1}-z)(E_{\g}-E_{\g_2}-
z)\cdots(E_{\g}-E_{\g_l}-z)} \eeq It is a kernal function in our
approach and expressions.

\section{}\label{AppendixD}

Furthermore, we can expand the partition function as the power
series of time $t$. \beq\label{ADPFone} \sum_\g A^{\g \g}=\sum_{\g}
\e^{-\I E{\g}t}\left[1+\sum_{n=1}^{\infty}\frac{(-\I
t)^n}{n!}C^\g_n\right]\eeq where the coefficient of $(-it)^n$ is as
below: \beqa C_n^{\g}&=&\sum_{m=n-1}^\infty\frac{n~
(-1)^{m-n+1}}{(m+1)(m-n+1)!}\left.\left[\Dif{z}{m-n+1}R_{\g}^{m+1}(z)\right]\right|_{z=0}
\eeqa It is easy to prove by expanding $\e^{\I zt}$ and noting the
derivative at $z=0$.

From the definition of partition function, we also have
\beq\label{ABPFfour} \sum_\g A^{\g \g}=\sum_\g\e^{-\I
\widetilde{E}_\g t}=\sum_\g\e^{-\I (E_\g+\Delta E_\g) t}=\sum_{\g}
\e^{-\I E{\g}t}\left[1+\sum_{n=1}^{\infty}\frac{(-\I
t)^n}{n!}(\Delta E_\g)^n\right]\eeq Here, we have rewritten the
Hamiltonian eigenvalues as \beq \widetilde{E}_\g=E_\g+\Delta
E_\g\eeq

After finishing the proof of the following relation
\beq\label{ADtask} C^\g_n=\left(C^\g_1\right)^n \eeq we can obtain
\beq \sum_\g\e^{-\I \widetilde{E}_\g t}=\sum_\g\e^{-\I
(E_\g+C_1^\gamma) t} \eeq Obviously \beq
C_1^\g=\sum_{m=0}\frac{(-1)^m}{(m+1)!}\left[\Dif{z}{m}R_{\g}^{m+1}(z)\right]\Bigg|_{z=0}\eeq

Now let us begin our task. It is clear that if
$C_{n+1}^{\g}=(C_1^{\g})\times C_n^{\g}$ is verified then
Eq.(\ref{ADtask}) is proved. From the definition of $C_n^\gamma$, we
see that, \beqa C_n^{\g} \times C_1^{\g}
    &=&\left\{\sum_{m_1=n-1}^\infty\frac{n(-1)^{m_1-n+1}}{(m_1+1)(m_1-n+1)!}
    \left.\left[\Dif{z}{m_1-n+1}R_{\g}^{m_1+1}(z)\right]\right|_{z=0}\right\}\nonumber\\
& & \times
    \left\{\sum_{m_2=0}^\infty\frac{(-1)^{m_2}}{(m_2+1)!}\left.\left[\Dif{z}{m_2}R_{\g}^{m_2+1}(z)\right]\right|_{z=0}\right\}
    \nonumber\\
&=&\sum_{M=n} (-1)^{M-n}\sum_{m=0}^{M-n}
\frac{n}{(M-m)(M-m-n)!(m+1)!}
\left.\left[\Dif{z}{M-n-m}R^{M-m}(z)\right]\left[\Dif{z}{m}R_\gamma^{m+1}(z)\right]\right|_{z=0}
\eeqa The last step has used the index transformations $M\rightarrow
m_1+m_2+1$ and $m_2\rightarrow m$. While, we also know that \beq
C_{n+1}^{\g}=\sum_{M=n}^\infty(-1)^{M-n}
\left.\frac{(n+1)}{(M+1)(M-n)!}\left[\Dif{z}{M-n}R_{\g}^{M+1}(z)\right]\right|_{z=0}\eeq

So, to prove Eq.(\ref{ADtask}), we should prove the equation below
(where $n\geq 1$).

\begin{equation}\label{ADtaskone}\Dif{z}{M-n}R_\g^{M+1}(z)=\sum_{m=0}^{M-n}\frac{n}{n+1}\frac{M+1}{M-m}
\frac{(M-n)!}{(m+1)!(M-n-m)!}\left(\Dif{z}{M-n-m}R_\g^{M-m}(z)\right)\left(\Dif{z}{m}R_\g^{m+1}(z)\right)
\end{equation}
Or for simplicity, set $M-n=k$
\begin{equation}\Dif{z}{k}R_\g^{k+n+1}(z)=\sum_{m=0}^{k}\frac{n}{n+1}\frac{k+n+1}{k+n-m}
\frac{k!}{(m+1)!(k-m)!}\left(\Dif{z}{k-m}R_\g^{k+n-m}(z)\right)\left(\Dif{z}{m}R_\g^{m+1}(z)\right)
\label{ADtasktwo}
\end{equation}
Note that it is not Leibnitz' rule for the $n$th derivative of a
product since the differential functions connect with the summation
index. It is clear that when $k=1$, we can directly verify
Eq.(\ref{ADtasktwo}) is valid. That is \beqa & &
\sum_{m=0}^{1}\frac{n}{n+1}\frac{1+n+1}{1+n-m}\frac{1}{(m+1)!}\frac{1}{(1-m)!}
\left(\Dif{z}{1-m}R^{1+n-m}(z)\right)\left(\Dif{z}{m}R_\g^{m+1}(z)\right)\nonumber\\
& & =(n+2)R^{n+1}(z)\dif{z}R_\g(z)=\dif{z}R_\g^{1+n+1}(z)\eeqa

Assume up to a given $l$ Eq.(\ref{ADtasktwo}) is correct, and then,
in the case of $l+1$, we can also prove it is correct as below.
After the one order differential, using the Leibnitz' rule for high
order derivative of a product, it follows that \beq
\Dif{z}{l+1}R_\g^{l+1+n+1}=(l+n+2)
\sum_{j=0}^{l}\frac{l!}{j!(l-j)!}\left(\Dif{z}{l-j}R_\g^{l+1}\right)\left[\Dif{z}{j}\left(R_\g^{n}\dif{z}R_\g\right)\right]\eeq
Since Eq.(\ref{ADtasktwo}) is assumed to be correct up to a given
$l$, Eq.(\ref{ADtaskone}) is valid up to $M=l+n$ $(n\geq 1)$. Thus
we can make use of Eq.(\ref{ADtaskone}) to replace the $(l-j)$th
derivative in above equation except for $j=0$. \beqa
\Dif{z}{l+1}R_\g^{l+1+n+1}&=&(l+n+2)\left\{\sum_{j=1}^{l}\frac{l!}{j!(l-j)!}
\left[\sum_{m=0}^{l-j}\left(\frac{j}{j+1}\right)\left(\frac{l+1}{l-m}\right)\frac{(l-j)!}{(m+1)!(l-j-m)!}\right.\right.\nonumber\\
& &\left.\left.
\times\left(\Dif{z}{l-j-m}R_\g^{l-m}\right)\left(\Dif{z}{m}R_\g^{m+1}\right)\right]\left[\Dif{z}{j}\left(R_\g^{n}\dif{z}R_\g\right)\right]+
\left(\Dif{z}{l}R_\g^{l+1}\right)\left(R_\g^{n}\dif{z}R_\g\right)\right\}\eeqa
Interchanging the summations over $j$ and $m$, we have
\beqa\label{ADtaskthree}
\Dif{z}{l+1}R_\g^{l+1+n+1}&=&\sum_{m=0}^{l-1}\frac{(l+n+2)}{(l-m)}\frac{(l+1)!}{(m+1)!}\left[\sum_{j=1}^{l-m}
\left(\frac{j}{j+1}\right)\frac{1}{j!(l-m-j)!}\left(\Dif{z}{l-j-m}R_\g^{l-m}\right)\right.\nonumber\\
& & \left.\times \Dif{z}{j}\left(R_\g^{n}\dif{z}R_\g\right)\right]
\left(\Dif{z}{m}R_\g^{m+1}\right)+\frac{l+n+2}{n+1}
\left(\Dif{z}{l}R_\g^{l+1}\right)\dif{z}R_\g^{n+1} \eeqa Specially,
the terms in the square bracket can be simplified as \beqa &
&\sum_{j=1}^{l-m} \left(\frac{j}{j+1}\right)\frac{1}{j!(l-m-j)!}
\left(\Dif{z}{l-j-m}R_\g^{l-m}\right)\Dif{z}{j}\left(R_\g^{n}\dif{z}R_\g\right)\nonumber\\
& & =\sum_{j=1}^{l-m} \frac{1}{j!(l-m-j)!}
\left(\Dif{z}{l-j-m}R_\g^{l-m}\right)\Dif{z}{j}\left(R_\g^{n}\dif{z}R_\g\right)\nonumber\\
& &\quad -\sum_{j=1}^{l-m}
\left(\frac{1}{j+1}\right)\frac{1}{j!(l-m-j)!}
\left(\Dif{z}{l-j-m}R_\g^{l-m}\right)\Dif{z}{j}\left(R_\g^{n}\dif{z}R_\g\right)\nonumber\\
& &=
\frac{1}{(l-m)!}\left[\frac{1}{(l+1+n-m)}\Dif{z}{l+1-m}\left(R_\g^{l+1+n-m}\right)
-\frac{1}{n+1}\left(\Dif{z}{l-m}R_\g^{l-m}\right)\dif{z}R_\g^{n+1}\right]\nonumber\\
& &-\frac{1}{n+1}\sum_{j=1}^{l-m}
\frac{1}{(j+1)}\frac{1}{j!(l-m-j)!}
\left(\Dif{z}{l-j-m}R_\g^{l-m}\right)\left[\Dif{z}{j+1}R_\g^{n+1}\right]\nonumber\\
& &=
\frac{1}{(l-m)!}\left[\frac{1}{(l+1+n-m)}\Dif{z}{l+1-m}\left(R_\g^{l+1+n-m}\right)
-\frac{1}{n+1}\left(\Dif{z}{l-m}R_\g^{l-m}\right)\dif{z}R_\g^{n+1}\right]\nonumber\\
& &-\frac{1}{(n+1)}\frac{1}{(l+1-m)!}\left[
\Dif{z}{l+1-m}R_\g^{l+1+n-m}-\left(\Dif{z}{l+1-m}R_\g^{l-m}\right)R_\g^{n+1}\right.\nonumber\\
& &\left.
-(l+1-m)\left(\Dif{z}{l-m}R_\g^{l-m}\right)\dif{z}R_\g^{n+1}\right]\nonumber\\
&=&\frac{n}{n+1}\frac{1}{(l-m)!}\frac{l-m}{(l+1-m)(l+1+n-m)}\Dif{z}{l+1-m}R_\g^{l+1+n-m}\nonumber\\
& &
+\frac{1}{n+1}\frac{1}{(l+1-m)!}\left(\Dif{z}{l+1-m}R_\g^{l-m}\right)R_\g^{n+1}\eeqa

Substituting it into Eq.(\ref{ADtaskthree}), we immediately arrive
at \beqa \label{ADtaskfive}\Dif{z}{l+1}R_\g^{l+1+n+1}&=&
\sum_{m=0}^{l-1}\frac{n}{n+1}\frac{l+1+n+1}{l+1+n-m}
\frac{(l+1)!}{(m+1)!(l+1-m)!}\left(\Dif{z}{l+1-m}R_\g^{l+1+n-m}(z)\right)\left(\Dif{z}{m}R_\g^{m+1}(z)\right)\nonumber\\
& &+\sum_{m=0}^{l-1}\frac{1}{n+1}
\frac{l+n+2}{l-m}\frac{(l+1)!}{(m+1)!(l+1-m)!}\left(\Dif{z}{l+1-m}R_\g^{l-m}\right)R_\g^{n+1}\left(\Dif{z}{m}R_\g^{m+1}(z)\right)\nonumber\\
& &+\frac{l+n+2}{n+1}
\left(\Dif{z}{l}R_\g^{l+1}\right)\dif{z}R_\g^{n+1}\nonumber \\
&=&
\sum_{m=0}^{l+1}\frac{n}{n+1}\frac{l+1+n+1}{l+1+n-m}
\frac{(l+1)!}{(m+1)!(l+1-m)!}\left(\Dif{z}{l+1-m}R_\g^{l+1+n-m}(z)\right)\left(\Dif{z}{m}R_\g^{m+1}(z)\right)\nonumber\\
&
&-\frac{l+n+2}{n+1}\left[\frac{1}{l+2}\left(\Dif{z}{l+1}R_\g^{l+2}\right)R_\g^n
-\frac{1}{n+1}\left(\Dif{z}{l}R_\g^{l+1}\right)\dif{z}R_\g^{n+1}\right.\nonumber\\
& &\left.-\sum_{m=0}^{l-1}
\frac{1}{l-m}\frac{(l+1)!}{(m+1)!(l+1-m)!}\left(\Dif{z}{l+1-m}R_\g^{l-m}\right)R_\g^{n+1}\left(\Dif{z}{m}R_\g^{m+1}\right)\right]
\eeqa Now, the problem is changed into the proof that the below
expression is zero. \beqa\label{ADtasksix} &
&\frac{1}{l+2}\left(\Dif{z}{l+1}R_\g^{l+2}\right)R_\g^n
+(n-1)\left(\Dif{z}{l}R_\g^{l+}\right)\dif{z}R_\g^{n+1}\nonumber\\
& &-\sum_{m=0}^{l-1}
\frac{1}{l-m}\frac{(l+1)!}{(m+1)!(l+1-m)!}\left(\Dif{z}{l+1-m}R_\g^{l-m}\right)R_\g^{n+1}\left(\Dif{z}{m}R_\g^{m+1}\right)\eeqa
Similar to deal with Eq.(\ref{ADtaskone}), that is, carrying out
derivation like Eq.(\ref{ADtaskfive}), we can obtain the $(l+1)$th
the derivative in the first term. \beqa \Dif{z}{l+1}R_\g^{l+2}&=&
\sum_{m=0}^{l-1}
\frac{l+2}{l-m}\frac{(l+1)!}{(m+1)!(l+1-m)!}\left(\Dif{z}{l+1-m}R_\g^{l-m}\right)R_\g\left(\Dif{z}{m}R_\g^{m+1}(z)\right)\nonumber\\
& &+(l+2) \left(\Dif{z}{l}R_\g^{l+1}\right)\dif{z}R_\g \eeqa
Substitute it into Eq.(\ref{ADtasksix}), we immediately see the
expression (\ref{ADtasksix}) is zero. It means that
(\ref{ADtasktwo}) is proved to be correct for $l+1$. Therefore, we
have proved Eq.(\ref{ADtask})

\end{document}